# AI Federalism:

# Shaping AI Policy within States in Germany

Anna Jobin [a],*, Licinia Güttel [a], Laura Liebig [a, b], Christian Katzenbach [a, b]

[a] Alexander von Humboldt Institut for Internet and Society, Französische Str. 9, D-10117 Berlin

[b] Zentrum für Medien-, Kommunikations- und Informationsforschung ZeMKI, Universität Bremen, Linzer Str. 4, D-28359 Bremen

* Corresponding author: anna.jobin@hiig.de

## Abstract

Recent AI governance research has focused heavily on the analysis of strategy papers and ethics guidelines for AI published by national governments and international bodies. Meanwhile, subnational institutions have also published documents on Artificial Intelligence, yet these have been largely absent from policy analyses. This is surprising because AI is connected to many policy areas, such as economic or research policy, where the competences are already distributed between the national and subnational level. To better understand the current dynamics of AI governance, it is essential to consider the context of policy making beyond the federal government. Although AI may be considered a new policy field, it is created, contested and ultimately shaped within existing political structures and dynamics. We therefore argue that more attention should be dedicated to subnational efforts to shape AI and present initial findings from our case study of Germany. Analyzing AI as a policy field on different levels of government will contribute to a better understanding of the developments and implementations of AI strategies in different national contexts.

# 1. Introduction

Discourses about Artificial Intelligence (AI) have raised concerns about risks but also potentials of implementing these technologies. Policy actors have been contributing to the salience of the issue by publishing policy documents on Artificial Intelligence. The focus of much of recent AI governance research situates and analyses AI strategies, white papers, ethics guidelines and similar documents issued by national governments and international institutions (e.g., Jobin et al., 2019, Schiff et al., 2021).

Subnational institutions have also published documents on Artificial Intelligence. However, these have been mostly overlooked by research on AI governance. This is surprising, because Artificial Intelligence is connected to many policy areas, such as economic or research policy, where the competences are already distributed between the national and the subnational level. Therefore, to better understand the current dynamics of AI governance, it is important to take into account the context of policy making beyond the federal government, and adopt a perspective we term 'AI federalism'. When, and on what grounds, do subnational policy institutions issue policy documents on artificial intelligence? Is AI policy treated as part of, or different from, digital policy on the subnational level? What are the links, or contrasts, to national policy efforts, and how can the subnational AI policy discourse be situated within existing federal structures? Answers will contribute to better understanding the influence of national AI strategies on the subnational policy level and vice versa, as well as the discours about and around artificial intelligence in AI governance more broadly.

In this paper, we therefore focus on subnational efforts to shape Artificial Intelligence. We first give a brief overview of current and related work in AI governance research. We then discuss the theoretical and political context of federalism and digital policy via the example of Germany. Subsequently, we present first results from our empirical analysis of subnational AI policy efforts by German states. Lastly, we discuss our results and relate them to the implications of accounting for AI federalism as an important dynamic in the governance of AI.

# 2. AI Governance and Policy Research

Overall, AI governance is still an unorganized field that is inhabited by multiple, diverse stakeholders (Butcher & Beridze, 2019, p. 89). It comprises various modes, initiatives, and implementations (Schmitt, 2021). For instance, according to Brundage and Bryson (2016), AI policy comprises three different sub-categories. For one, there are what they call direct AI policies, which are policies that have been issued specifically for AI-based technologies. Secondly, there are indirect AI policy, i.e. policies implemented for other technologies but indirectly affecting AI-based systems too. Lastly, the authors list what they call AI-relevant policies, by which they mean policies from 'other' domains that would benefit from knowledge about AI-based technologies (Brundage & Bryson, 2016, p. 5). Companies and other private sector organizations occupy a central role in AI governance to the point that they sometimes appear to be better positioned for leading AI governance than governments (Dafoe, 2018, p. 125). Another important perspective centers on inclusion and participatory policy-making -- or lack thereof (e.g. McKelvy & Macdonald, 2019).

The focus of this paper lies on policy documents issued by public sector actors on AI. The publication of such policy documents highlights the vital role of the state, notably in collaborations with research and the private sectors in the domains of AI developments and implementation (Ulnicane et al., 2020, p. 166). This approach implies that AI is considered a strategic issue as nation states "anticipate a significant impact on the global distribution of economic, military, and political power." (Fischer & Wenger, 2021, p. 172). This competitive perspective as 'a new space race' (Ulnicane et al. (2021, p. 80), which can be clearly distinguished from a vision that centers on cooperative alliances between national governments (cf. also Smuha, 2021, on competition vs. convergence of AI regulation).

In some cases, national strategies point to ethical principles, industrialization, and research, while neglecting AI regulation because of uncertainties with regards to managing these new technologies (Radu, 2021, pp. 9-11). Other recent work analyses frames and narratives in national AI policy in the broader sense (e.g. Bareis and Katzenbach, 2021; Radu, 2021; Köstler & Osewaarde, 2021; Roberts et al., 2021). While most of these rhetorics and controversies were identified in national policy discourses of more economically developed countries, which account to a large amount of technological progress, countries of the Global South are just starting to become part of the discussion. For instance, Filgueiras (2021, pp. 13-14) analyzed national AI strategies of Latin American countries from a policy design perspective and uncovered the importance and delimiting nature of governance modes, hierarchy, and populism.

Nevertheless, governments do not operate in a vacuum but within existing governance structures. Specific cases highlight notably the role of subnational structures, especially within countries that have instituted federal governments. For example in the US, particularly direct AI policies were implemented by states in the form of laws governing specific contingencies of self-driving vehicles (Brundage & Bryson, 2016, p. 5). Further legislative approaches primarily target algorithmic systems, including New York's 'algorithmic transparency bill' (for the city's use of algorithms), Florida's 'iBudget' system (personalized services for persons with disabilities), and Massachusetts' ban on the government's facial recognition usage (Lucero, 2020). US cities, like San Francisco for instance, passed their own legislation too and are a key site of AI implementation, but are not often considered to be part of the AI discussion (Schmitt, 2021). A plausible reason for this could be their isolation from the global level (p. 9). Although being more isolated than the federal government, subnational governments can offer advantageous solutions to regulatory issues and problems related to implementation. Squitieri (2021, p. 148) emphasizes the benefit of states to implement, for instance, a 'wrong' regulatory regime first. On a national level, a supposedly destructive choice could lead to difficulties to reverse such decisions and could therefore entail down-stream effects. Moreover, state governments have the capacity to adapt their standards and norms to localized circumstances (p. 152). In the same vein, cities and communities have been proposed as important vectors for innovating governance of AI (Verhulst et al., 2021).

Germany's political multilevel system is also a prime example to demonstrate the ways in which AI is entangled within existing federal structures. For one, this is due to Germany's position in the European Union, but secondly, and central for this article, it is so because of the federal system within the nation state. The distribution of competencies across ministries and state agencies on different governmental levels -- particularly prominent in economic and research policies -- does indeed make AI governance endeavours more complex. Besides the fact that German states (Bavaria and Hessen) have already implemented ministries for digitization on a subnational level while the national government lags behind (Deutscher Bundestag, 2021), subnational institutions have also published policy documents on AI. Beyond the impact these documents may or may not have on the current research landscape, they provide a foundation for how AI is implemented and negotiated at a level 'closer' to the citizens. If the solution to better AI governance is, indeed, 'AI localism' (Verhulst et al., 2021), then the perspective of AI federalism deserves more attention.

## 3. German Federalism and Digital Policy

Federalism and digital policies

Federalism is an institutional design that divides the power between at least two distinct levels of government to guarantee a balance of power and more regional autonomy, with a high degree of multi-level decision making, cooperative federalism and decentralization in the German case (Härtel, 2017, pp. 199-200; Scholta et al., 2019, pp. 3273-3275). At the same time, the sharing of power between the federal and states' level comes with some caveats, because the allocation of power and responsibilities over policy areas is not always clear. This can incentivize politicians to take credit or shift responsibilities for policy issues however it best suits their interests. The blend of areas of responsibility can make it harder for citizens to identify the responsible level of government for a certain issue and thus further motivate regional politicians to take credit for issues dealt with at the federal level (Gross & Krauss, 2021, pp. 3-4).

Federalism and the development of digital policies are closely intertwined, especially in the German case. Digitization is a so-called cross-cutting issue that touches on many policy areas, such as research, education, economy or public infrastructure. The interconnection between federalist structures and the shaping of digital policies stems from the shared competences between the federal and states' level, as well as by the fact that the states execute the laws made at the federal level. Digitization is a relatively new policy field linked to several policy areas, which is why the shaping and creation of digital policies exacerbates German federalism's characteristics.

Digital Policies in Germany

Germany has a long tradition of federalism where special emphasis is put on the protection of fundamental rights through a federal power division (Härtel, 2017, pp. 199-200). Germany's federal institutions influence the shaping of policies, as joint-decision making is an integral part of its institutional setup and competences are not always clearly distributed between the two levels of government. Germany is a particularly interesting example to study federalism and its effect on policies because the 16 German states are deemed to have considerable leeway in policy-making. This applies especially to the domains where states have exclusive power, such as basic education, police, cultural policy, and media policies.

The states are jointly responsible with the federal state for taxation, social and labor policies, as well as for executing the laws made at the federal level (Gross & Krauss, 2021, p. 5).

It has previously been argued that Germany is a good example to study the variation of digital policies on the states' level due to the fact that German states have considerable decision-making power -- and, thus, the autonomy to draft "independent" digital policies -- while sharing the same institutional and cultural background (Siewert & König, 2019, p. 254). To understand how digital policies are created, it is necessary to capture the developments at the federal as well as the states' level because digital policy simultaneously involves both levels. States co-shape digitization policies, which is probably best illustrated by the fact that all German states have released digital strategies by 2017. Some states have even created digital ministries. Interesting in this regard is also the role of political parties at the states' level. Siewert and König (2019) found that parties address digitization most often in states where digital change has reached high levels, but is neither saturated nor in its beginning (p. 260).

Additionally, the establishment of digital policies does not only involve the relation between the federal level and the states' level, but also the coordination and competitions between states themselves. The German multi-level governance system and cooperative federalism may induce elements of regional competition, especially on the inter-regional level (Benz, 2007). Such a policy competition is not characterized by economic competition during which governments behave like market actors but rather by states' governments searching for approval for their decisions (p. 426). This argument can be applied to states' behavior during the drafting of digital policies, as they might use these policies to improve or maintain their socioeconomic standing relative to other states (Siewert & König, 2019, p. 254).

Because of joint decision-making, German federalism is sometimes seen as hindering innovation and digitization. Regarding the drafting and implementation of e-government policies, some scholars describe that the slow process to determine the responsible government level can lead to suboptimal decisions because, once a decision is reached, it might be based on the lowest common denominator principle due to the number of actors involved (Scholta et al., 2019, pp. 3273-3275). Others, however, argue that federalism as an institutional framework designed to balance powers and guarantee civil liberties has the potential to find the best regulations and approaches to deal with digital innovations, while promoting these values (Härtel, 2017). A related argument positions federalist structures as being able to produce more innovative projects and policies on the subnational level because local politicians may be exposed to less risk or aiming for a higher position, thereby challenging the status quo (Rose-Ackerman, 1980, p. 614). An analysis of digital strategies has shown that the states' digital strategies go beyond the federal government's expectations. As states' digital strategies are shaped by a mix of coordination, cooperation, competence, and competition between actors (Härtel, 2017, p. 214), their drafting procedure and content depends on factors such as party competition, traditional economic structures, financial strength, regional references as well as social conditions. Despite these differences, some convergence regarding objectives and impetus can be observed (Härtel, 2017, p. 203).

In short, the example of Germany's federal structures demonstrate that these modes of decision-making impact how digital policies are drafted. This applies to digital policies in particular because they touch many policy areas and, thus, many levels of government. Yet, how federal structures impact individual digital policy areas can vary, which is why we analysed empirical data of AI policy documents that exist at the states' level.

## 4. German Subnational AI Policy Documents

We illustrate the usefulness of AI federalism, i.e. a change of perspective, through a preliminary mapping of German AI policy documents on the subnational level. Our empirical results reveal that all 16 federated states have indeed developed AI or digital strategies, as well as other policy documents referring to the advancement and governance of AI technologies.

### Data Collection

The AI policy documents were collected between March 2021 and September 2021. We define policy documents as documents addressing policy issues mentioning Artificial Intelligence. As there is no generally agreed definition for Artificial Intelligence, we included each document that uses the term "Artificial intelligence" without applying a definition of our own to include or exclude specific technologies. To identify relevant documents, the search terms "Künstliche Intelligenz" (Artificial Intelligence) and "KI" (AI) were used on the regional governments' central websites. For each of the 16 German Länder (states), at least one document was included in the selection. For those Länder in which more than one document appeared, only documents in which either AI is mentioned several times, or in which a specific section is dedicated to AI, were included.

In total we retrieved 40 AI policy documents on the subnational level. Three of these documents were labelled as national and subnational at the same time, as they were issued by joint federal and states' conferences, composed by all 16 state ministers and a representative of the federal government. Overall, the documents included in our dataset were published between 2016 and 2021. Almost every document was authored by the public sector, except for four documents which were (co-)authored by private sector actors and one document authored by a research actor. Within the public sector, state governments and state chancelleries are the main issuers.

## Data Analysis

For each state, we found at least one document covering AI, but the extent to which AI is covered varies considerably. For instance in Mecklenburg-Western Pomerania and Saarland, AI is not covered extensively. The Länder that published the most documents in our dataset are Baden-Württemberg, Hessen and Northrine-Westfalia, with four AI related documents in each state. This seems to indicate that states with a low population size and less economic power tend to publish fewer documents.

While some states (e.g., Baden-Württemberg, Hessen, Thuringia) published digital strategies between 2016 and 2018, the keyword Artificial Intelligence was mentioned rarely. Only after the release of the national German AI strategy in October 2018, the first AI strategies were published at the Länder level. Baden-Württemberg was the first to publish their AI strategy, namely in early November 2018. Most states published their AI strategies in 2019 or 2020. Saxonia released its AI strategy in September 2021 and represents the most recent AI strategy in our dataset. Some states have not yet published strategies exclusively dedicated to AI.

The AI policy documents vary regarding their mode of drafting. In some states, an assessment or public recommendation by a private sector actor (e.g., a consultancy agency or a health insurance company) or a research actor preceded the AI strategy, as in Hessen, Berlin/Brandenburg, and Mecklenburg-Western Pomerania. Many strategies mention public-private-partnerships in their strategies.

We found that several terms and rhetorics which were part of the national AI strategy and other national policy documents have also been adopted by subnational actors. For instance, the state government of Baden-Württemberg echoed the national strategy by formulating the goal of creating an 'ecosystem' for AI. In this context, they mention their so-called 'Cyber Valley' with several economically powerful companies such as Amazon, BMW, Bosch, or Daimler (Landesregierung Baden-Württemberg, 2018, p. 1). This reference to Silicon Valley, as well as the strong focus on research and further development, exemplifies the competitive character of the strategy. Interestingly, Baden-Württemberg, a subnational state, mentions its commitment to collaborate with France and French initiatives several times in its AI strategy (Landesregierung Baden-Württemberg, 2018, pp. 2-5).

Similarly, the 'Hightech Agenda', published by the Bavarian state government, promises to expand Bavaria's position at the forefront of technology research. This also includes, among other things, a comparison with competitor states, and the development of Munich into a 'world-class AI centre' (Bayerische Staatsregierung, 2019, p. 7). Northrine-Westphalia also claims such a leading role for itself (Landesregierung Nordrhein-Westfalen, 2018).

In contrast, the joint efforts of the states Berlin and Brandenburg are characterised on the one hand by the fact that initially more of an inventory on the topic of AI was carried out and published (Technologie Stiftung Berlin, 2018). On the other hand, the innovation strategy 'innoBB 2025' that followed from that only mentioned AI with regards to the media and creative branch and Internet of Things technologies (Senat Berlin & Landesregierung Brandenburg, 2019, p. 16).

Ethical principles are also part of some subnational strategies. Bremen's position paper, for example, highlights seven ethical standards including 'priority of human action and supervision', 'privacy and data quality management', 'transparency', and 'diversity, non-discrimination and fairness' (Der Senator für Wirtschaft, Arbeit und Häfen & Die Senatorin für Wissenschaft, Gesundheit und Verbraucherschutz, 2019, p. 5).

Moreover, the digital strategy of the federal state of Hamburg references the findings of the Data Ethics Commission that emphasized the importance of high standards in AI development to prevent discrimination and secure legal certainty. In addition to promoting small and medium-sized enterprises, Hamburg is focusing on the use of AI in administration, e.g. through a chatbot (Senatskanzlei – Amt für IT und Digitalisierung, 2020, p. 57).

Narratively, some documents also give an assessment of whom they are targeting. The Hessian AI strategy states over several pages which citizens are affected by specific AI technologies on a regular basis. Among them are pupils, small business owners, nurses, and managing directors from all different backgrounds and cities within Hessen (Hessische Ministerin für Digitale Strategie und Entwicklung, 2021, p. 6-12). This example shows the citizen-centric orientation of some of the strategies and is accentuated through the first goal of the Hessian strategy, namely to give citizens the chance to use their quality of life and opportunities for personal development through digitization (Hessische Ministerin für Digitale Strategie und Entwicklung, 2021, p. 16). In Northrine-Westfalia and Saxonia, formats of citizen consultation prior or accompanying the release of the policy documents have been launched. The Saxon AI strategy, published only very recently, mentions that citizens need to be able to trust AI in relation to the citizen consultation process (Staatsregierung Sachsen, 2021, p. 32) and announces its goal that every citizen as a consumer needs to have some knowledge on AI (p. 39).

Overall, specific fields of application are repeatedly pointed out in the subnational AI strategies. Often, these fields of application match the states' regional economic identity, with e.g., Baden-Württemberg mentioning its automobile manufacturers (Landesregierung Baden-Württemberg, 2018, p. 1) and Hessen Frankfurt as a finance hub (Hessische Landesregierung, 2016, pp. 3-4). This might show that states use their AI policy to promote their economic position relative to other states. In addition to administration, business, and research, these include, for instance, schools, hospitals as well as sustainability and climate change. For the latter point, the strategy of Schleswig-Holstein can be used as an example, which aims to promote intelligent electricity grids and mobility concepts (Staatskanzlei Schleswig-Holstein, 2019, p. 11).

In sum, our preliminary analysis reveals that states are very invested in shaping AI policies at the subnational level. They have adapted some topics of the national strategy and linked the development of AI policies to their own regional economic identity. The issues and topics addressed thus remain varied. Additionally, the AI policy documents vary regarding their relationship to the public, and to citizen involvement more generally.

## 5. Discussion and outlook

Our empirical analysis offers some preliminary insight into the subnational landscape of AI policy documents, demonstrating that AI governance research benefits from including the perspective of AI federalism. Our mapping of AI policy documents calls for a more thorough analysis of the relation between national and subnational policy initiatives. While our analysis is not exhaustive, our preliminary findings and interpretations based on the documents in our dataset already point to future strands of research that could be fruitful. Regarding our case study Germany, the shared cultural and institutional context and the possibilities to draft different strategies in different states at the same time have indeed proven a coherent context to analyze different AI policy documents, similar to analyses of digital party programs and digital strategies (Siewert & König, 2019, p. 254). The fact that each state addresses Artificial Intelligence as a policy issue in some ways raises interesting questions. One such question would be whether the states that address AI the most share common characteristics. Another salient point might be to ask whether there are specific incentives for states to engage in AI policy. Our research also raises the question whether the publication of AI strategies correlates with the level of digital change reached in a state, or can even be linked to the presence of a digital ministry.

As for states' common characteristics that address AI more or less relative to other states: We found that West German states with rather large population sizes address AI more than some East German states. Still, it is unclear how the publication of AI policies is related to states' economic power, especially considering certain industries' economic power that might have more interest to signal a strong commitment to AI.

Regarding the question whether AI is part of subnational digital policies: a possible answer could be similar to the motivation behind more general digital policy documents, in which case, the main driver would be intra-state competition (Siewert & König, 2019, p. 254). If states aim to promote their position relative to other states, AI policy documents and related initiatives are a means for states to maintain their economic standing and signal to other actors their involvement in AI techniques. Another reason for states to develop AI policy documents is the fact that, because of the different levels of responsibility in federal systems, voters might hold regional politicians accountable for federal policies (Gross & Krauss, 2021, p. 4). This might motivate regional politicians to signal to voters that they care about the new policy issue AI and develop their own documents.

Because AI is often seen as a new technique requiring new policies (Djeffal, 2020, p. 278), it is interesting to compare the development of AI policy documents at the subnational level with the emergence of other documents addressing digital innovations. Scholta et al. (2019) showed that e-government policies were developed very slowly on the subnational level, as the complex federal structures hindered quick progress because of the difficulty to assign areas of responsibility. However, we found the opposite regarding AI strategies: German states were very quick to publish AI strategies after the release of the national strategy. This shows that federalism is capable of producing quick outcomes in the context of AI. However, it also challenges whether learnings from federalism and digital policy apply to artificial intelligence, and underlines the need to study AI federalism more in-depth. A possible explanation for the faster pace of AI policy on the federal level might be found in the fewer resources required to draft these strategies. But perhaps, it might be explained by a sense of urgency by state governments to publish AI strategies as "prestige objects" in terms of innovation.

Our dataset contains several digital strategies that have been previously analyzed by Härtel (2017), because they specifically mention Artificial Intelligence. Perhaps unsurprisingly, we find that Härtel's argument on the merit of studying coordination, cooperation, competence, and competition in relation to the drafting and content of such policy documents also applies to our corpus. Especially relevant in the realm of AI policy documents seems Härtel's point regarding the need to analyse the mode of governance in light of the actors' mode of interaction. The coordination of states in relation to each other and to the national government in the process of drafting AI strategies deserve more attention. Future research could, for example, investigate narrative frames within subnational AI strategies in comparison to the national strategy.

Based on our preliminary evidence we contend that AI research would benefit from a stronger focus on the subnational level. Aside from Germany, many other countries are organised in similar federal structures, which could also make them suitable as objects of research and comparison. Such a focus on the subnational level echoes the explicit call by the OECD for cooperation between different levels of government with regard to digital policy in general (Mello & Ter-Minassian, 2020, p. 16). Therefore, it is evident that analyzing AI as a policy field on different levels of government will contribute to a better understanding of AI governance more broadly.

*AI Federalism: Shaping AI Policy within States in Germany* 15Radu, R. (2021). Steering the Governance of Artificial Intelligence: National Strategies in Perspective. *Policy and Society, 40*(2), 178–93.

Roberts, H. et al. (2021). Safeguarding European Values with Digital Sovereignty: An Analysis of Statements and Policies. *Internet Policy Review* 10(3). https://policyreview.info/pdf/policyreview-2021-3-1575.pdf

Rose-Ackerman, S. (1980). Risk Taking and Reelection: Does Federalism Promote Innovation? *The Journal of Legal Studies*, *9*(3), 593–616. https://doi.org/10.1086/467654

Schiff, D., J. Borenstein, J. Biddle, & K. Laas. (2021). AI Ethics in the Public, Private, and NGO Sectors: A Review of a Global Document Collection. *IEEE Transactions on Technology and Society* 2(1), 31–42.

Schmitt, L. (2021). Mapping global AI governance: A nascent regime in a fragmented landscape. *AI and Ethics*. https://doi.org/10.1007/s43681-021-00083-y

Senat Berlin, & Landesregierung Brandenburg (2019). innoBB 2025. Gemeinsame Innovationsstrategie der Länder Berlin und Brandenburg. https://innobb.de/sites/default/files/2020-01/inno_bb_2025_a4-broschuere_final_download_0.pdf

Scholta, H., Niemann, M., Halsbenning, S., Räckers, M., & Becker, J. (2019). Fast and Federal—Policies for Next-Generation Federalism in Germany. *Hawaii International Conference on System Sciences 2019 (HICSS-52)*. https://aisel.aisnet.org/hicss-52/dg/policies_for_digital_government/5

Siewert, M. B., & König, P. D. (2019). On digital front-runners and late-comers: Analyzing issue competition over digitization in German subnational elections. *European Political Science Review*, *11*(2), 247–265. https://doi.org/10.1017/S1755773919000109

Senatskanzlei – Amt für IT und Digitalisierung (2020). Digitalstrategie für Hamburg. Retrieved from: https://www.hamburg.de/contentblob/13508768/703cff94b7cc86a2a12815e52835accf/data/download-digitalstrategie-2020.pdf

Smuha, N. A. (2021). From a 'Race to AI' to a 'Race to AI Regulation': Regulatory Competition for Artificial Intelligence. *Law, Innovation and Technology, 13*(1), 57–84.

Squitieri, C. (2021). Federalism in the Algorithmic Age. *Duke Law & Technology Review,* 139-158.

Staatskanzlei Schleswig-Holstein (2019). Künstliche Intelligenz. Strategische Ziele und Handlungsfelder für Schleswig-Holstein. Retrieved from: